\documentclass[12pt,twoside, a4paper]{article}
\usepackage{color}
\usepackage{graphicx}
\usepackage{authblk}
\usepackage{url}
\usepackage{float}
\usepackage{subcaption}
\usepackage[export]{adjustbox}
\usepackage{amsmath}

\begin{document}

\title{Modelling the Daily Electricity Demand of Electric Vessels in Plymouth}
\author[1]{Lauren Ansell}
\affil[1]{School of Engineering, Computing and Mathematics,
Faculty of Science and Engineering, Plymouth University, Drake Circus, Plymouth, PL4 8AA, UK}

\maketitle

\section{Abstract}

The international Maritime Organization (IMO) has set the target of reducing the emissions from the shipping sector to at least 50\% of the 2008 levels. One potential method to cut emissions is to convert vessels to battery powered propulsion in a similar manner to that which has been adopted for motor vehicles. Although, battery powered propulsion will not be suitable for all vessels, the conversion of those that are will lead to an increase in the energy demand from the national grid. This study uses historic port call data is used to model the timings of arrivals and the number of vessels in the port of Plymouth to predict the increase in additional energy demand required for battery powered vessels through a period of 24 hours as a greater proportion of the fleet move to battery powered propulsion.\\
\linebreak
\textbf{Keywords:} electric vessels, energy demand, big data, clean maritime, real-world vessel data

\section{Introduction}

As we move towards the decarbonisation of travel and the reduction in the use of fossil fuels, much of the focus of policy previously has been on the motor vehicle sector \cite{Reducingemissions,TakingCharge}, however this focus is now shifting towards transport in the maritime sector. In 2019, the Department of Transport published the Clean Maritime report \cite{cleansea} stating the aim to reach zero emissions shipping by 2050 to align with the International Maritime Organization (IMO) target of reducing the emissions in international shipping by at least 50\% when compared to the levels of 2008 \cite{IMO}. To address the reduction in emissions from vessels, similar technology employed in the electrification of motor transport can be employed in marine transport. In April 2021, the city of Plymouth became the home of the UK’s first sea-going electric ferry \cite{eboat} and the first UK city to install a network of electric charge points for battery propulsion vessels \cite{echarge}.\\

As potentially more vessels more towards battery powered propulsion, this implies that a greater draw of energy will be required from the National grid. Understanding when and how much additional energy could be necessary is key to making sure the country has robust energy resources in place to handle the extra demand. Another issue which could also arise, along with ensuring the supply exists to meet the demand, is verify that the required energy can be delivered form the nearest substation at the times it will be required without detriment to the existing users.\\

Currently we have little to no data regarding the charging demand from electric vessel, with the technolnogy being new to the sector. However the demand of energy from the charging of electric vehicles has been extensively modelled using a variety of methods. The demand for electricity in the Greater Stuttgart area of Germany was studied by \cite{MALLIG} by using the mobiTopp model to simulate the demand and journeys of electric vehicles and hence derive the power demand for this on a weekly basis. The study by \cite{Arias} took a different approach and combined the use of historical traffic and weather data to identify overarching patterns of travel through the application of machine learning techniques. The time at which electric vehicle charging began was treated as a random variable to identify charging demand patterns during the day through probabilistic methods. Van Viliey et. al.\cite{VANVLIET} investigated the current and future impact electric cars has on electricity demand and the infrastructure required for generation and distribution by way of efficiency, cost and emissions of greenhouse gases. Studies have carried out to assess both the impact on energy demand and the reduction in emissions, for example the study by \cite{DIAS}, which used a bottom-up approach to the modelling, uses 2005 as the base year to predict the emissions reduction and energy demand in 2035 in S\~ao Paulo and into the comparison of different modes of electric transportation to assess the contrast in demand, for example \cite{GRENIER} in Christchurch, New Zealand. \\

In this paper, the approach we will take to model the daily electricity demand will be a simplification of that used in \cite{Arias}, where the arrival patterns for vessels that commonly frequent the port of Plymouth are derived from previous port calls and the charge demand is calculated using the average number of arrivals. \\

The remainder of the paper is organised as follows. Section 3 describes the data sources used in the analysis; Section 4 reports the results of the analysis; finally, concluding remarks are presented in Section 5.

\section{Data and Methodology}
\begin{figure}[H]
\begin{center}
\includegraphics[scale=0.4]{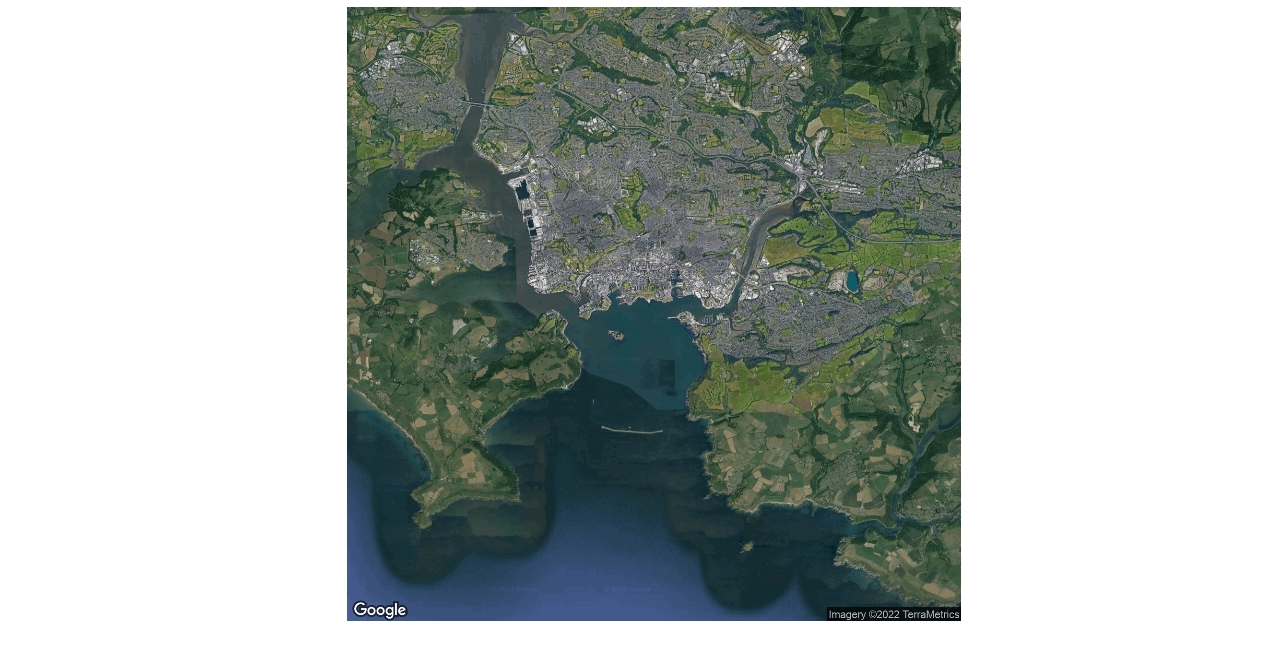}
\caption{Satellite image of the port of Plymouth \cite{google}}
\label{fig:port}
\end{center}
\end{figure}

Plymouth is an international sea port on the south west coast of the UK which covers an area of approximately 4620 square kilometres, shown in figure \ref{fig:port}. The city has recently seen the installation of 3 charge points in-situ and fully operational with charging powers of 25kW, 75kW and 150kW \cite{echarge}, the last of these being a world first.\\

To construct the predictions of the energy demand from charging electric vessels, historical data of the port calls for Plymouth covering the time period from the 1\textsuperscript{st} January 2019 to the 31\textsuperscript{st} December 2019 were used. This time period was selected as it was not affected by any restrictions put in place in response to the Covid-19 pandemic in March 2020 or changes in regulations following the UK exit from the European Union following the transition period on 1\textsuperscript{st} January 2021. The data was supplied by Fleetmon \cite{fleetmon}, where a port call is defined as a vessel travelling into the polygons of the Plymouth port area. The original dataset contained information on 88,379 individual calls from June 2009 to November 2021.\\

The vessels that call at the port of Plymouth range from 2 meters in size, for vessels usch as pleasure crafts, to 587 meters in size for the Naval vessels. For this analysis, only vessels under 25 meters in length were considered as this length of vessel is more suitable for conversion to battery powered propulsion in regards to the power required from the motor to generate motion. The dataset was initially cleaned to remove any port calls which fell outside the time period of interest, after which vessels of length 25 meters or more where removed.  After the removal of these vessels, the dataset was reduced to 9,469 individual port calls and identified 31 different types of vessel. Many of these vessel types had a small frequency of calls and therefore vessel types with a frequency of port calls less than 500 were then removed so to capture the demand which would occur from the most common vessels accessing the port. This final stage of filtering reduced the types of vessels to five, shown in table \ref{table:vesselNos} below, and are the primary focus in this analysis:

\begin{table}[H]
\centering
\begin{tabular}{|l|l|}
\hline
Vessel type & Frequency\\
\hline
Sailing ship & 2056 \\
Fishing vessel & 1655  \\
Pusher/Tug & 1546\\
Yacht & 1176  \\
Trawler & 553 \\
\hline
\end{tabular}
\caption{Types of vessels under 25m recorded in Plymouth}
\label{table:vesselNos}
\end{table}

\begin{figure}[H]
    \centering
      \begin{subfigure}{0.48\textwidth}
        \includegraphics[width=\textwidth]{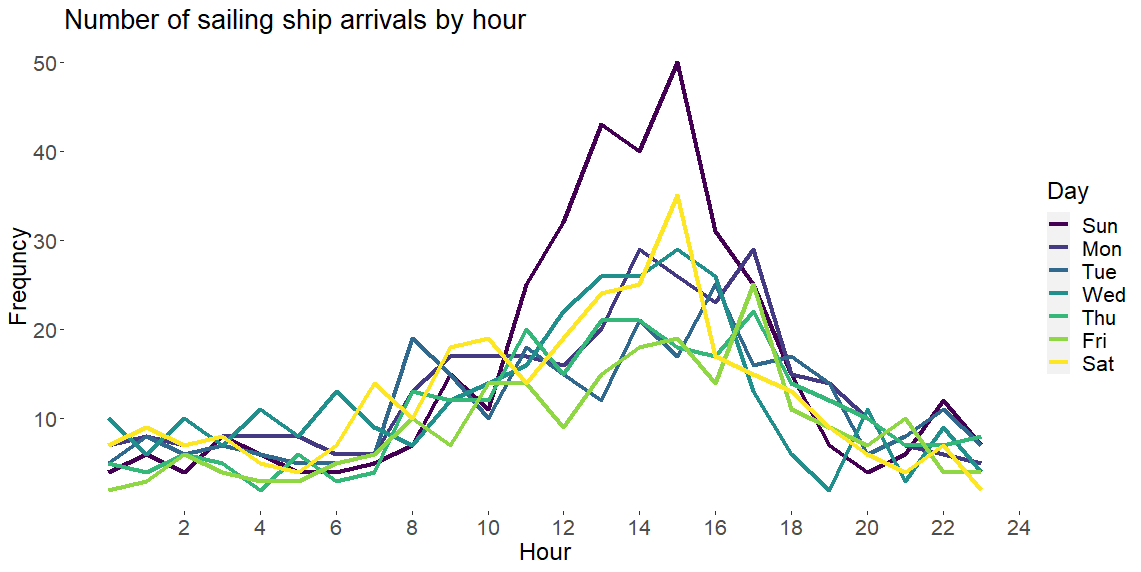}
          \caption{Sailing ship}
          \label{fig:patha} 
      \end{subfigure}
      \begin{subfigure}{0.48\textwidth}
        \includegraphics[width=\textwidth]{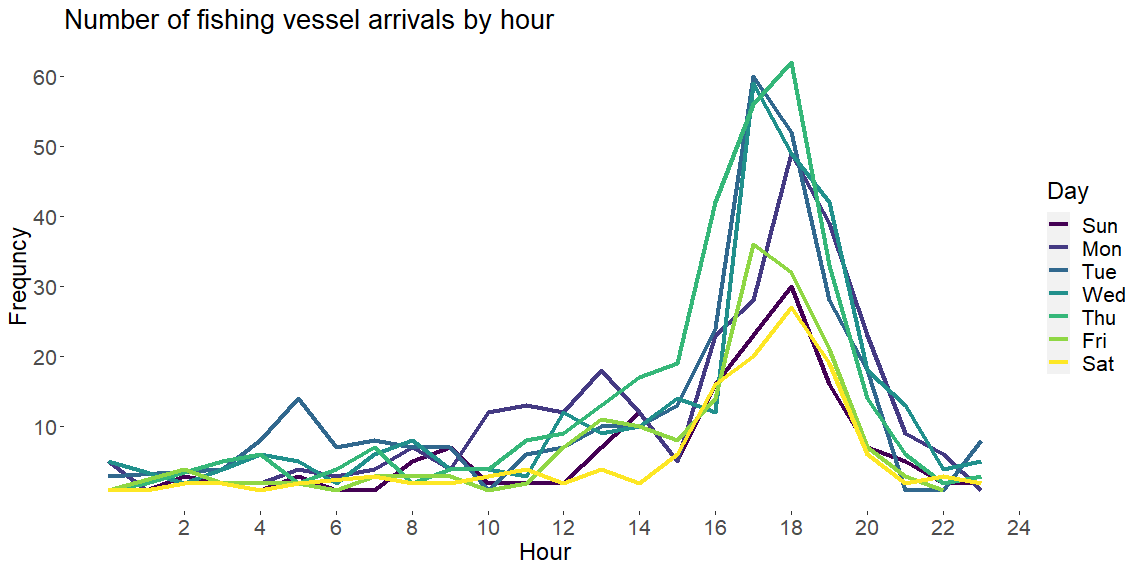}
          \caption{Fishing vessels}
          \label{fig:pathb} 
      \end{subfigure}
      \begin{subfigure}{0.48\textwidth}
        \includegraphics[width=\textwidth]{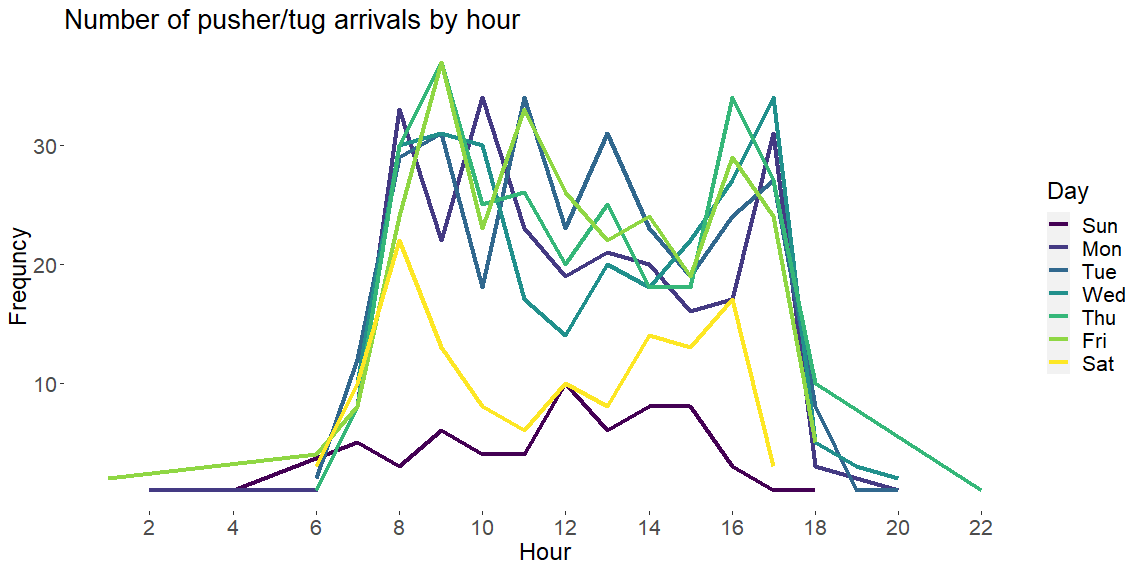}
          \caption{Pushers/Tugs}
          \label{fig:pathc} 
      \end{subfigure}
      \begin{subfigure}{0.48\textwidth}
        \includegraphics[width=\textwidth]{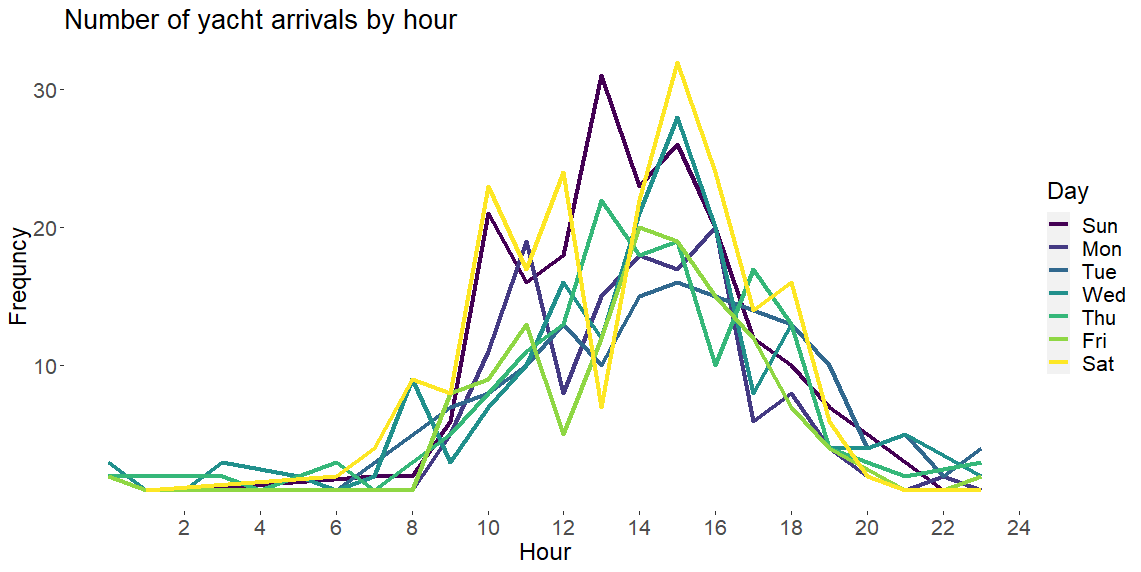}
          \caption{Yachts}
          \label{fig:pathd} 
      \end{subfigure}
            \begin{subfigure}{0.48\textwidth}
        \includegraphics[width=\textwidth]{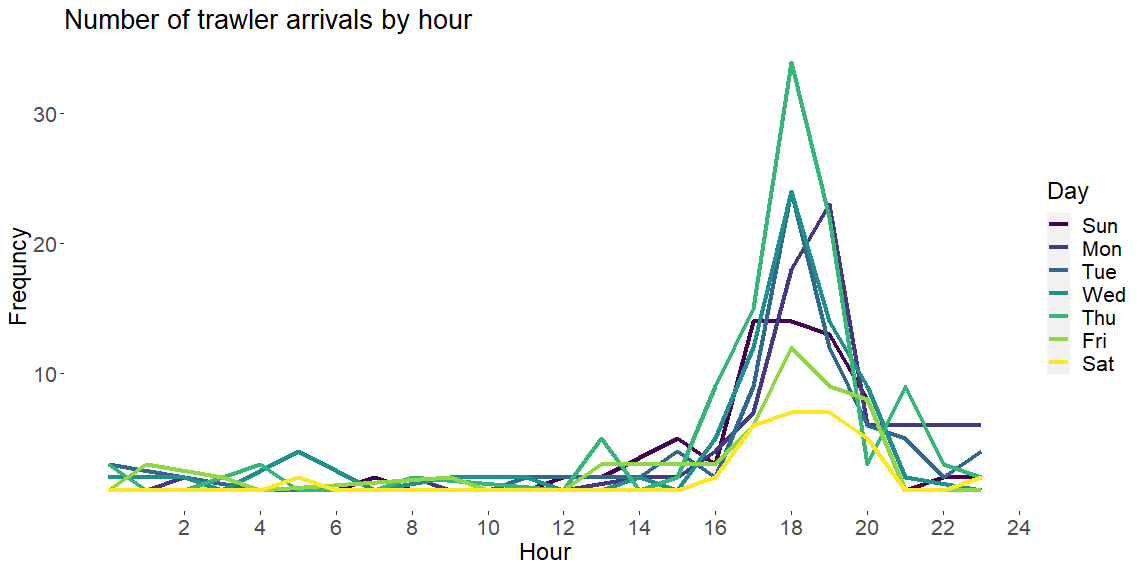}
          \caption{Trawlers}
          \label{fig:pathe} 
      \end{subfigure}
\caption{Number of arrivals by vessel type}
\label{fig:arrivals}
\end{figure}

The resulting dataset shows an acceptable balance between commercial vessels and leisure vessels, with the former comprising 53\% of the dataset, which allows for predictions from both leisure and commercial usage to be made. Once the vessel types have been selected for analysis,the lubridate \cite{lubridate} package within R \cite{R} was employed to extract the day of the week and the hour of arrival for each of the different vessel types to analyse the distribution throughout the 24 hour period of a day and are shown in figure \ref{fig:arrivals}. \\

\subsection{Commerical vessels}

We can see from the figure above that the fishing vessels, figure \ref{fig:pathb}, and trawlers, figure \ref{fig:pathe}, display an approximately similar pattern of behaviour over a period of 24 hours. We see small fluctuations in arrivals before a clear peak of arrivals beginning at approaching 4pm through until 6 pm. This pattern of behaviour is replicated on each day of the week, however with a reduction of almost half in the number of arrivals over the weekend period. Due to this pattern in the behaviour, we will assume that a trawler or fishing vessel will leave the port with the battery charged at 100\% which provides enough power until the vessel returns to port at the end of the working day to recharge.\\

Similar to the trawlers and fishing vessels, the pushers/tugs record lower numbers of arrivals over the weekend period. However, during the working week, we can see that arrivals in the port begin at 6am and continue on throughout the day, with the number of arrivals fluctuating until approximately 6pm. For these vessels, we will assume that each vessel will have a fully charged battery at the commencement of the working period, however, these vessels will ``top up'' their charge throughout the day upon returning to port by rapid charging. \\

\subsection{Leisure vessels}

As we saw a similar pattern across two of the types of commercial vessels, we find that the yachts and sailing ships present similar patterns in arrivals for leisure vessels. Both of these types of vessels see small fluctuations in arrivals throughout the day with the number of arrivals increasing from midday for yachts and 2pm for the sailing ships until 4pm. We note that the leisure vessels display the opposite behaviour to the work vessels with the highest number of port calls for these types of vessels being recorded on the weekend. For these vessels, we will assume that the vessel leaves the port with the battery at 100\% and will commence recharging the battery when the vessel returns to the port to recharge fully.\\

\subsection{Calculating the demand}

Once the patterns in behaviour have been identified, the potential power demand per hour will be calculated. To do this, a piecewise linear charging profile has been adopted of the form:
\begin{equation}
P(t)=
    \begin{cases}
        P_r & \quad 0<t \leq t_1\\
        P_r\left(\frac{t_2 - t}{t_2 - t_1}\right) & \quad t_1<t\leq t_2
    \end{cases}
\end{equation}
This is where $P_r$ is the charging power, this is determined by the charging mode and $t_1$ and $t_2$ are the times that determine the variation in the charging power magnitude as the battery nears full charge \cite{chargeprofile}. The total power demand is calculated for all the vessels by evaluating the following summation \cite{Arias},
\begin{equation}
P_t=\sum_{i=1}^n P_d (t).
\end{equation}
This results in the total hourly power demand being given by the sum of all the charging power being demanded by vessels at the particular time $t$. For the modelling it has been assumed that the $P_r$ value for the slow charging scenario is 75kW and the for the rapid charging scenario, we have $P_r=150$ kW. We have made the assumption for the slow charging scenario, the total time taken to achieve full charge is  hours and the vessel will begin charging in the hour at which it arrives back into port. In the rapid charging scenario, once again the assumption is made that charging begins within the hour the vessel arrives back into port and it takes one hour to achieve full charge.

\begin{figure}[h]
    \centering
      \begin{subfigure}{0.48\textwidth}
        \includegraphics[width=\textwidth]{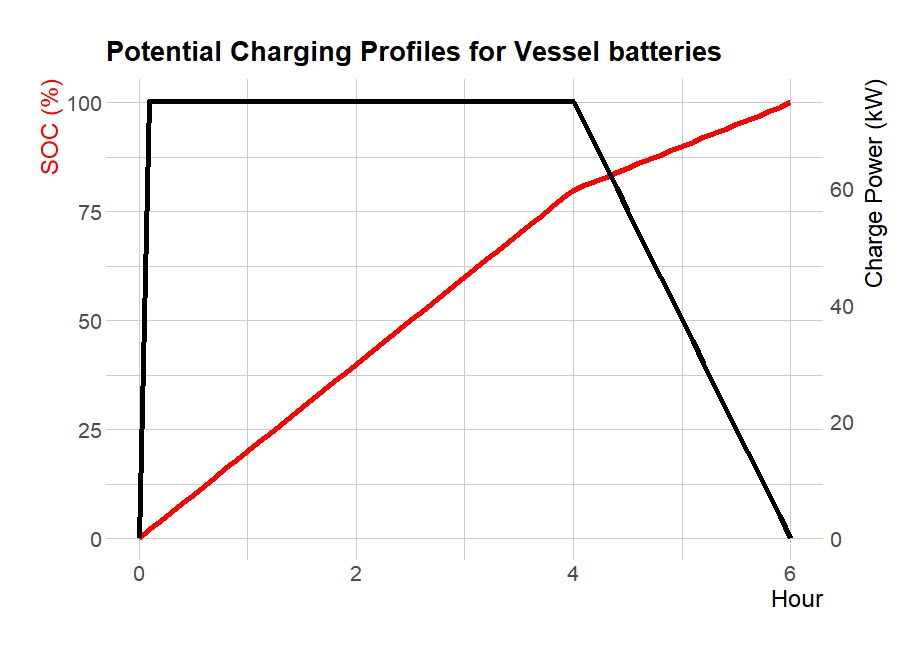}
          \caption{Slow charging}
          \label{fig:chargea} 
      \end{subfigure}
      \begin{subfigure}{0.48\textwidth}
        \includegraphics[width=\textwidth]{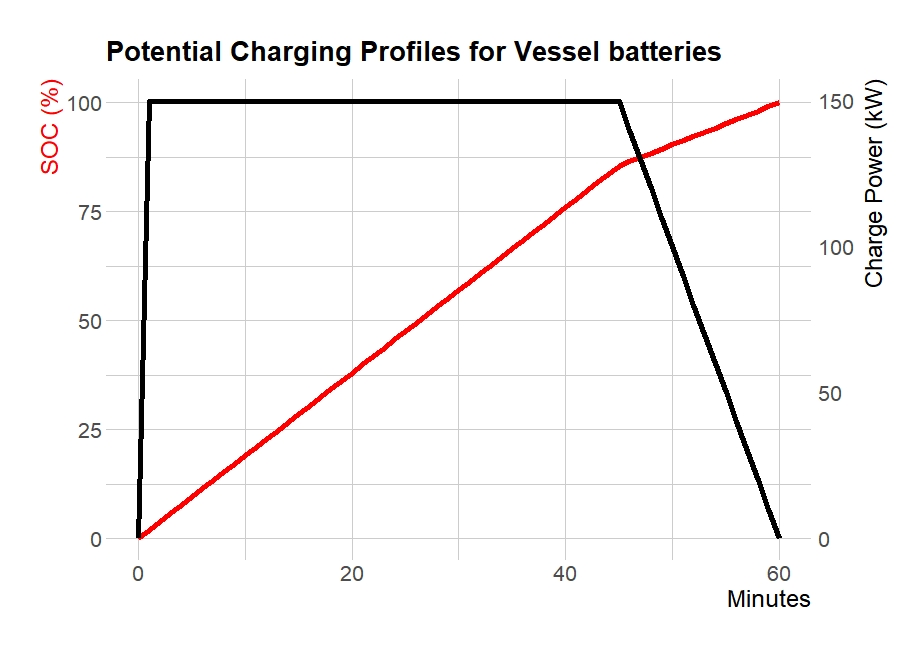}
          \caption{Rapid charging}
          \label{fig:chargeb} 
      \end{subfigure}
\caption{ Battery charge profiles for slow and rapid charging}
\label{fig:chargeprofile}
\end{figure}

As the distribution of arrivals remains consistent across a 24 hour period, the average of the arrivals in the hour interval has been calculated to model the number of arrivals on any given day and in turn used to calculate the power demand from charging.

\section{Result Analysis}

\begin{figure}[H]
    \centering
      \begin{subfigure}{0.48\textwidth}
        \includegraphics[width=\textwidth]{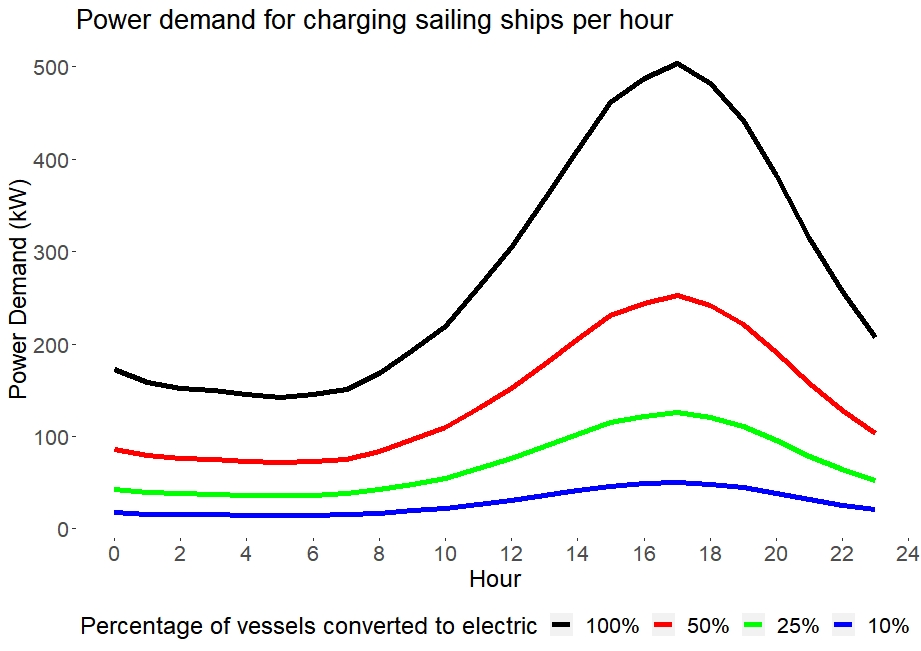}
          \caption{Sailing ship}
          \label{fig:demanda} 
      \end{subfigure}
      \begin{subfigure}{0.48\textwidth}
        \includegraphics[width=\textwidth]{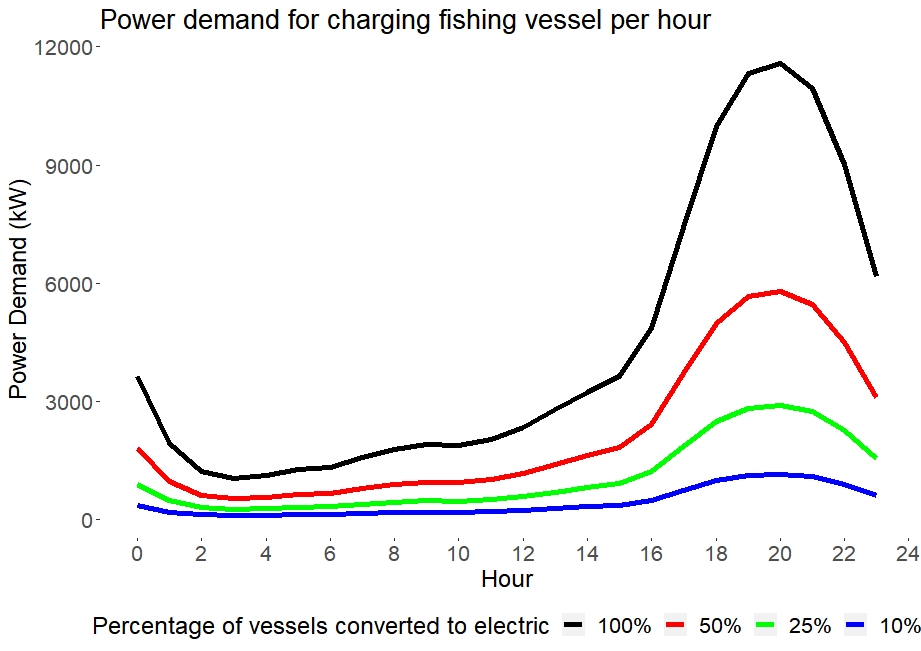}
          \caption{Fishing vessels}
          \label{fig:demandb} 
      \end{subfigure}
      \begin{subfigure}{0.48\textwidth}
        \includegraphics[width=\textwidth]{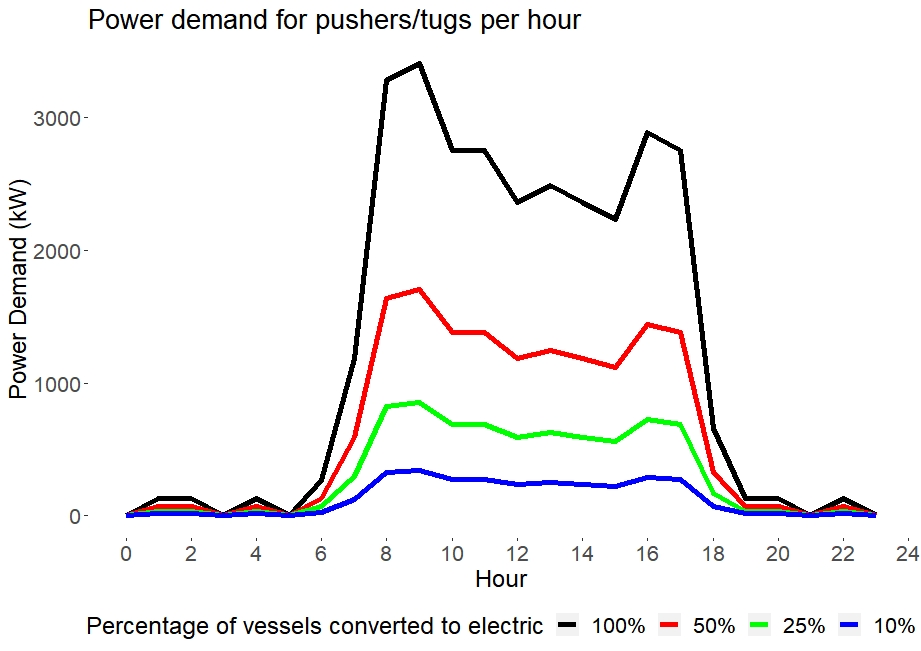}
          \caption{Pushers/Tugs}
          \label{fig:demandc} 
      \end{subfigure}
      \begin{subfigure}{0.48\textwidth}
        \includegraphics[width=\textwidth]{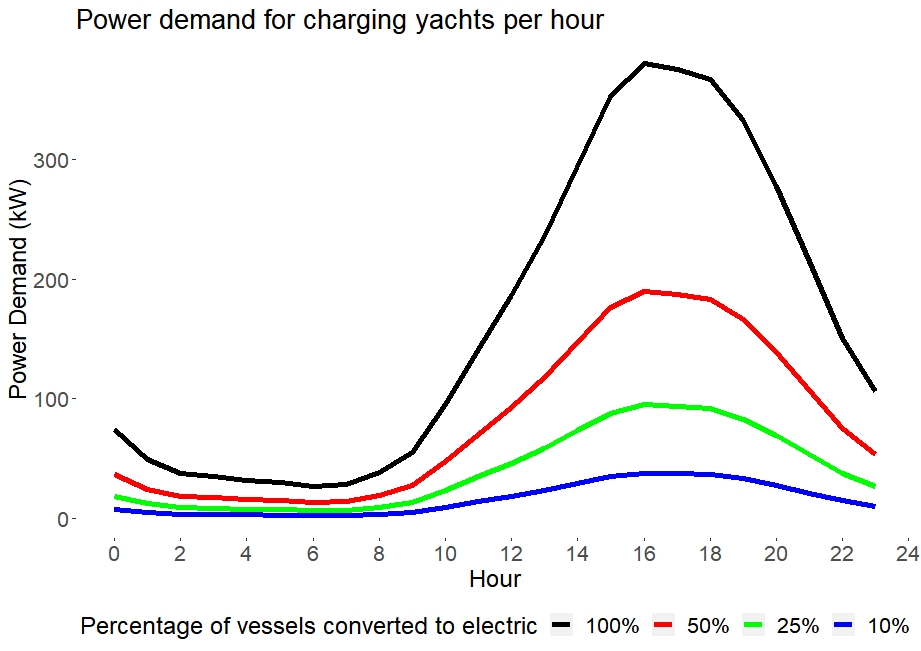}
          \caption{Yachts}
          \label{fig:demandd} 
      \end{subfigure}
            \begin{subfigure}{0.48\textwidth}
        \includegraphics[width=\textwidth]{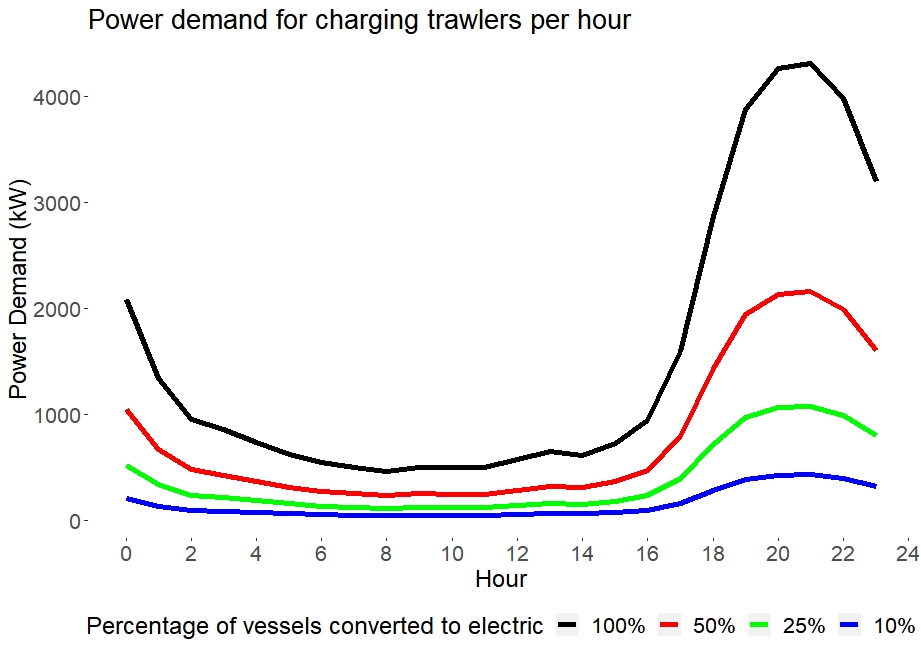}
          \caption{Trawlers}
          \label{fig:demande} 
      \end{subfigure}
\caption{Potential electricty demand  by vessel type}
\label{fig:demand}
\end{figure}

As seen with electric vehicles, there will be a gradual increase in the number of electric vessels and therefore to reflect this the total amount of charging power demand has been calculated on the condition that 10\%, 25\%, 50\% and 100\% of the vessels under analysis here switched to battery propulsion respectively, as shown in figure \ref{fig:demand}.

We that if at first only a small percent from each vessel type transfer to battery powered propulsion, initially there will be a modest amount of additional demand on the national grid. However this demand for maritime charging will potential occur where the demand for charging increases coincides with demand from other sources such as the charging of electric vehicles following the evening commute. The additive effect of the additional demand resulting from electrification of maritime vessel could potentially lead to creating a surge in demand in a short space of time.\\

As we move towards the 2050 deadline for clean maritime \cite{cleansea}, with the entire vessel fleet moving to battery power propulsion, we see the potential demand increase to a point where there is no point during a 24 hour period where energy is not being demanded due to the volatile nature of vessel arrivals. We also see a potential peak demand of power over 18 MW in the evening period should all the vessels convert to electric.

\section{Discussion}

The approach taken in this paper has been very simplistic, only taking into account the average frequency of the different types of vessel arrivals in the port of Plymouth in each hour and using set proportions of these averages to predict the potential electricity demand from the charging of the vessels. To create a more accurate model, further analysis should be done to investigate how vessel movements could possibly be tied to the weather conditions and seasonality, for example adopting a similar approach such as the one taken in \cite{Arias}. Here machine learning techniques such as clustering could be employed to identify patterns and seasonal variations in the vessel movement which would allow for a greater selection of vessel types to be included in the analysis and therefore producing a more robust prediction of the energy demand.  \\

Although it has been assumed here that there will not be a sudden switch to electric vessels in the main, the numbers converting to battery power will gradually increase over time and the analysis here has not taken into account any growth in the number of vessels at Plymouth that would potentially also be electric. Additional assumptions used in this analysis include the assumption that charging will begin once the vessel returns to port and that all the marine vessels in the port will be using batteries of the same power. Due to the tiny number of vessels using this technology we currently have no evidence to support such assumptions and further work should include treating the charging start time as a random variable, particularly in the case of the pushers and tugs where it is unlikely that charging will be required every time they return to the port but is not fully known at this point in time.\\

Due to the small number of vessel currently using this technology, there is limited data regarding the run distance and time of a single charge, which may also be affected by the weather conditions in which the vessel is sailing. This could result in an increase in the number of port calls as vessels will be required to return to port and recharge more frequently then the current distances and times recorded from traditional combustion engines. In addition to this, due to the technology being in it's infancy, rapid charging for marine vessels is not available yet, although discussed here as the preferred method of charging for vessels which would be required to make multiple journeys in a 24 hour period, akin to electric buses.  \\

In the future, as more vessels adopt electric motors, the data gathered can be used to test the assumptions made here more robustly and updated to reflect the realisation of fully a electric maritime system.\\

\section{Acknowledgements}

This work was supported through the Clean Maritime Call, a Maritime Research and Innovation UK (MarRI-UK) initiative supported by the Department for Transport, grant number 10008313.

  \bibliographystyle{abbrv}
  \bibliography{refs}

\end{document}